# Coexistence of Two Types of Liquid Structures at Platinum–Water Interfaces

Yitong Li[1,2], Qian Ai[1,2], Lalith Krishna Samanth Bonagiri[2,3], and Yingjie Zhang[1,2,4]*

1. Department of Materials Science and Engineering, University of Illinois, Urbana, Illinois 61801, United States

2. Materials Research Laboratory, University of Illinois, Urbana, Illinois 61801, United States

3. Department of Mechanical Science and Engineering, University of Illinois, Urbana, Illinois 61801, United States

4. Beckman Institute for Advanced Science and Technology, University of Illinois, Urbana, Illinois 61801, United States

*Correspondence to: yjz@illinois.edu

**Abstract:** Platinum–water interfaces underpin many electrochemical energy conversion processes. However, despite decades of research, the real-space liquid structure of these interfaces remains elusive. Using three-dimensional atomic force microscopy (3D-AFM), we mapped Pt–water interface in real space with angstrom-level resolution. Topographic imaging revealed atomically flat (type I) and stripe-like (type II) surface nanodomains. Force–distance profiles above type I domains exhibited oscillatory decay patterns with periodicity of ~0.33 nm, consistent with water. In contrast, type II domains showed stronger oscillations with larger periodicity of ~0.45 nm and extended decay lengths, indicative of a different liquid structure with stronger correlation and ordering. Wide-angle X-ray scattering (WAXS) measurements of pure water and a series of liquid n-alkanes revealed peaks at ~0.31 nm and ~0.46 nm, in agreement with 3D-AFM observations of type I and type II structures, respectively. Our findings uncover the coexistence of two types of liquid structures at Pt–water interfaces modulated by surface heterogeneity, enabling new understandings and design principles for energy conversion applications.

**TOC Graphic**

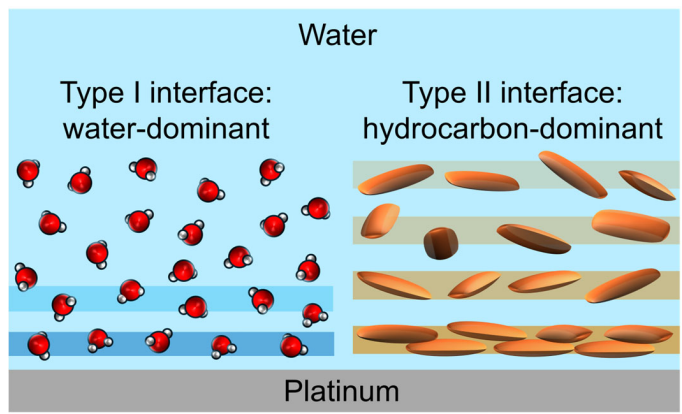

# 1. Introduction

Solid–liquid interfaces are important for a large range of real-world processes, including water purification,[1] thermal transport,[2–4] biological interactions,[5–8] energy storage,[9–13] and electrocatalysis.[14–19] In aqueous environment, interfacial water usually spans up to ~1 nm above the solid as a transition region between the solid and bulk water.[20,21] The interfacial water plays a crucial role in mediating many electrocatalysis processes such as hydrogen evolution reactions (HER),[22–24] oxygen reduction reaction (ORR),[25–27] and carbon dioxide reduction reaction ($CO_2$RR).[28,29]

The structure of interfacial water is known to be inhomogeneous. It has been demonstrated that the overall interfacial water density at hydrophilic mineral surfaces exhibits oscillatory decay patterns perpendicular to the solid surface, with a periodicity of ~3 Å, corresponding to the average intermolecular distance of hydrogen-bonded liquid water.[30,31] However, over the past decade, there have been multiple reports revealing that the interfacial liquid structure can be dramatically different when hydrophobic surfaces are immersed in aqueous solutions.[32–37] Hydrophobic materials, such as graphene and boron nitride, can attract hydrophobes of various origins onto their surfaces, giving rise to stripe-like morphology.[32,38,39] When immersed in water, these interfaces tend to gradually develop hydrocarbon-like liquid structure, with a periodicity larger than 3 Å.[32–36,40] However, at the intermediate regime where the solid surface is weakly hydrophilic or weakly hydrophobic, the solid–water interfacial structure remains unknown.

Among the systems with weak hydrophilicity/hydrophobicity, platinum–water interfaces stand out due to their high technological importance, particularly for electrochemical energy conversion applications. Pt exhibits high catalytic activity in aqueous environments, notably in ORR[41] and HER.[42] Interfacial water strongly modulates the Pt-catalyzed reactions. During alkaline HER, for example, the interfacial water coordination structure mediates the activation energy for water dissociation.[43] On the other hand, for HER in acidic electrolytes, it was demonstrated that the orientation of interfacial water molecules is responsible for the proton transfer kinetics.[44] Similarly, it was also reported that interfacial water and hydrophobic species strongly modulate ORR activities at Pt surfaces.[45]

The structure of Pt–water interfaces has been studied by a few techniques. Water contact angle measurements revealed that Pt surface is by nature hydrophilic, while adsorption or contaminations of unknown origin render the surface weakly hydrophobic.[46,47] Spectroscopic techniques, including X-ray absorption spectroscopy (XAS) and surface-enhanced Raman spectroscopy (SERS), have also been used to probe Pt–water interfacial structure.[43,44,48] However, these studies have focused on the spectroscopic features of water, and omitted the possibility that airborne contaminations can exist at the interface.

In this work, we used 3D-AFM to investigate the Pt–water interfacial structure. In the past decade, 3D-AFM has been widely used to image the real-space interfacial liquid density distribution at many solid surfaces, including minerals,[49–53] graphite/graphene,[33,36,54] transition metal dichalcogenides,[33,55,56] hexagonal boron nitride,[32] and self-assembled monolayers.[37] However, to the best of our knowledge, 3D-AFM has not been used to study metal–water interfaces. Here we used 3D-AFM to probe both the surface morphology of Pt and the surrounding liquid density distribution and compared the results with WAXS measurements of bulk liquids. After immersing a clean Pt in water, we observed two types of surface domains: flat morphology corresponding to clean Pt and periodic stripes attributed to airborne hydrocarbons. We found that the intrinsic Pt surface areas were surrounded by water layers, while stripy domains were covered by multiple hydrocarbon liquid layers.

## 2. Experimental Section

### 2.1. Materials

Pt(111) single-crystal disk substrate (purity > 99.999%, 12 mm diameter, 1 mm thick) was purchased from Princeton Scientific Corporation with one side polished to roughness < 0.01 micron and orientation accuracy < 1°. Milli-Q water (18.2 MΩ·cm) was obtained from Synergy UV water purification system (MilliporeSigma). n-Heptane (purity ≥ 99%, HPLC) was purchased from Sigma-Aldrich; n-octane (purity > 97.0% (GC)), n-nonane (purity > 98.0% (GC)), n-decane (purity > 99.0% (GC)), n-undecane (purity > 99.0% (GC)), n-dodecane (purity > 99.0% (GC)), n-tridecane (purity > 99.0% (GC)) and n-tetradecane (purity > 99.0% (GC)) were acquired from TCI America. Acetone (Fisher Chemical, purity ≥ 99.5%) and isopropanol (IPA, Fisher Chemical, purity ≥ 99.5%) were used for cleaning processes.

### 2.2. Pt Substrate Processing

Pt(111) single crystal was first cleaned with acetone and IPA: sonicated in each for 1–2 minutes using an ultrasonicator (Branson 2800, US Ultrasonics) and immediately blow-dried with $N_2$ (Airgas, Inc., ultra-high purity). The crystal was then treated with oxygen-plasma cleaner (Harrick Plasma PDC-32G Cleaner) at low radio frequency (RF) power (6.8 W) for 5 minutes to remove any remaining organic contaminations on the surface. Subsequently, the Pt substrate was annealed at 900 °C for 30 minutes (ramping rate: 10 °C/min until 900 °C, followed by free cooling) in 97% Ar / 3% $H_2$ (Airgas, Inc., ultra-high purity) gas environment at a total pressure of around 1.92 Torr. After cooling to room temperature, the Pt substrate was sealed in a glass vial (pre-cleaned using acetone and IPA) and then carried to an AFM instrument for imaging in either air or water.

### 2.3. In-Air AFM Surface Topography Mapping

All in-air x-y topography imaging were performed on a Cypher S AFM (Asylum Research, Oxford Instruments) and with Tap300Al-G probe (BudgetSensors). The previously processed Pt substrate was glued to a sample disk with double-sided Kapton tape. Surface imaging was performed in alternating current (AC), amplitude modulation mode. The AFM cantilever parameters and key imaging parameters are shown in Tables S1 and S2, respectively. The cantilever parameters were obtained using the same cantilever calibration processes as described in our previous publication.[55]

## 2.4. Liquid-Phase AFM Experiments

All liquid-phase AFM measurements were performed on a Cypher ES Environmental AFM using FS-1500AuD probe (Asylum Research, Oxford Instruments). The experimental protocol was similar to those described in our previous publications.[55,57–59] Before carrying out the experiment, the probe was soaked first in acetone for 30 minutes, in IPA overnight, and in Milli-Q water for 30 minutes. Then the probe was treated with UV Ozone for 5 minutes. The Pt substrate (after cleaning) was assembled into the AFM liquid cell, after which 100 μL of fresh Milli-Q water was dropped into the assembled cell using a pipette (Eppendorf Research Plus) so that the Pt substrate surface was immersed in water. The cell chamber was then purged and sealed in argon gas. Liquid-phase AFM measurements were conducted in AC, amplitude modulation mode. The AFM cantilever metrics and imaging parameters are shown in Tables S1 and S2, respectively. Before each x-y topography mapping or each 3D imaging, the cantilever was driven at its resonance frequency at a few microns above the Pt(111) substrate. During 3D-AFM imaging, the cantilever phase and amplitude were recorded as a function of spatial coordinates x, y, and z.

## 2.5. 3D-AFM Data Processing

Initially, adjacent phase-z and amplitude-z curves were averaged to ensure statistical significance. The $z = 0$ position was defined as the point with approximately half an interlayer distance to the left side of the first phase minimum. These averaged phase and amplitude curves were then calibrated such that the $z = 0$ positions were aligned, and phase values at the farthest z positions approached 90°. Force-z curves were subsequently obtained by conservative force reconstruction from the aligned amplitude and phase curves, using the algorithms developed in Ref. 60. A bi-exponential background was fitted to each curve and subtracted. Subsequently, a series of background-subtracted force curves were averaged again, and the averaged force curve was fitted using an oscillatory decay formula for parameter extraction.

## 2.6. X-ray Scattering Measurements

The X-ray scattering experiment was conducted on a home-built instrument (with the support of Forvis Technologies, Santa Barbara). This equipment was composed of Xenocs Genix3D ULD microfocus Cu Kα X-ray source (0.15418 nm) and high-speed Dectris Pilatus 300 detector (172 microns pixel size, 500 eV resolution), enabling 2D data acquisition. Using silver behenate powder as a calibration standard, the sample-to-detector distance was determined to be 0.14 m. Before loading the liquid samples, the empty quartz capillary (Hampton Research Corporation, outside diameter 1.5 mm) was first measured as the background signal. We then used a syringe to inject liquid into the capillary while ensuring that air bubbles were absent. The capillary was sealed by tape, and then we obtained the total signal of the liquid-containing capillary. The data acquisition time of both empty capillary and sample was 10 min for all n-alkanes and 30 min for water. The 2D data were integrated azimuthally using Fit2D software and the intensity was plotted as a function of momentum transfer (Q) within the range of ~3 nm$^{-1}$ to ~22 nm$^{-1}$. The scattering signal of the samples was obtained by subtracting the background signal from the total signal.

### 2.7. X-ray Photoelectron Spectroscopy Measurement

X-ray photoelectron spectroscopy (XPS) was performed using a Kratos AXIS Supra$^+$ spectrometer (Kratos Analytical) with a monochromatic Al Kα (1486.69 eV) X-ray source. The same Pt sample in the previous AFM experiments was pre-treated with argon plasma (NPC 3000, Nano-Master Incorporation), first at RF power 100 W for 1800 s, then at 200 W for 300 s. The pressure of argon plasma treatment was 200 mTorr. The treated sample was immediately transferred to the XPS chamber for measurements. All XPS data were processed using CasaXPS. The spectra were calibrated to the adventitious carbon 1s peak at 284.8 eV.[61,62]

## 3. Results And Discussion

### 3.1. Pt Surface Structure

We first used AC amplitude modulation mode AFM to measure Pt surface morphology both in air and in pure water after annealing in reducing environment. Surface step height measurements and XPS confirmed that the prepared substrate exhibited Pt(111) surface orientation and had a valence state of zero (Figure S1). In air, the Pt surface exhibited a uniform, atomically flat morphology with a root mean square roughness (Rq) of 31 pm (Figure 1a), considerably smaller than Pt(111) monoatomic step height of ~238 pm.[63–65] When immersed in water, however, two distinct surface morphologies emerged: atomically flat regions (type I, Rq ~ 20–45 pm) similar to those observed in air, and regions with periodic stripe-like features (type II, Rq ~ 25–60 pm). Figure 1b presents an area where both morphologies coexist, with type II characterized by periodic stripes oriented diagonally in the lower right part. Morphologies of both types from different locations are shown in Figure 1c,d, confirming consistent features. Quantitative height profiles (Figure 1e) highlight relatively smooth and random height fluctuations for type I domains, whereas type II domains exhibit periodic height variations with a peak-to-peak amplitude of 80–150 pm and a periodicity of 4–5 nm. A gallery of x-y topography maps and height profiles are shown in Figure S2, revealing the extensive presence of both type I and type II areas at Pt–water interfaces. Some stripe-like features underwent morphology changes after continuous scan within the same area (Figure S3), signifying that these species were likely weakly physisorbed.

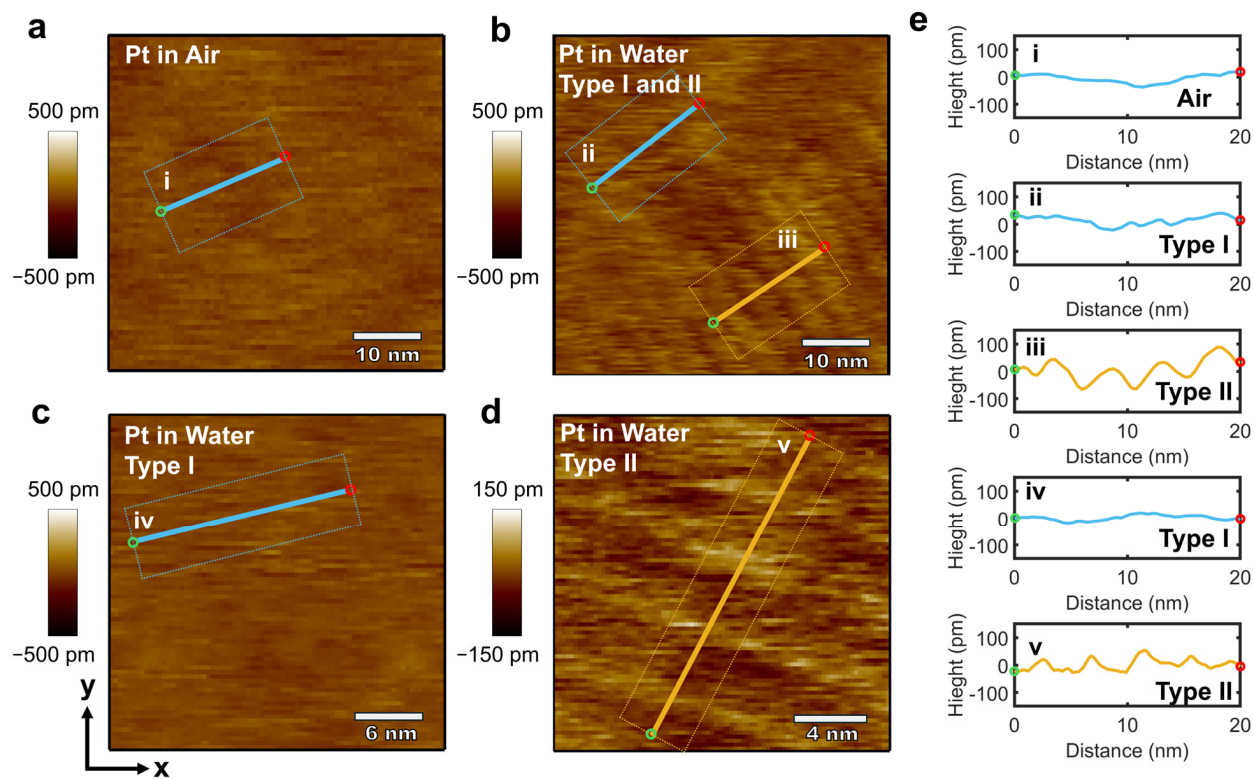

**Figure 1**. AFM surface height maps (x-y plane) of Pt in air and in water. (a) Pt surface in air. (b–d) Different spots of Pt surface in water, revealing the coexistence of type I (atomically flat) and type II (stripy) areas (b), solely type I area (c), and solely type II area (d). (e) Height profiles along the solid lines from (a–d), marked by the corresponding labels i–v. Each height profile in (e) was produced from the data in the corresponding dashed box in (a–d), by averaging the data perpendicular to the solid line and plotting the averaged height vs distance along the solid line. Green and red dots denote the starting and ending points, respectively, for both the solid lines in (a–d) and the height profiles in (e).

## 3.2. Local Liquid Structure at Pt–Water Interfaces

After imaging the substrate surface topography in liquid, we switched to 3D imaging mode to measure the liquid structure above type I and type II surface domains. As shown in Figure 2a,b and Figures S4–S6, we observed ubiquitous phase oscillations in both types of areas, although type I area exhibited weaker oscillations with smaller periodicity. By averaging multiple x-z phase map slices along x direction, we obtained one curve in Figure 2c,d. A series of these averaged phase curves are presented in Figure 2c,d, revealing consistent layered patterns.

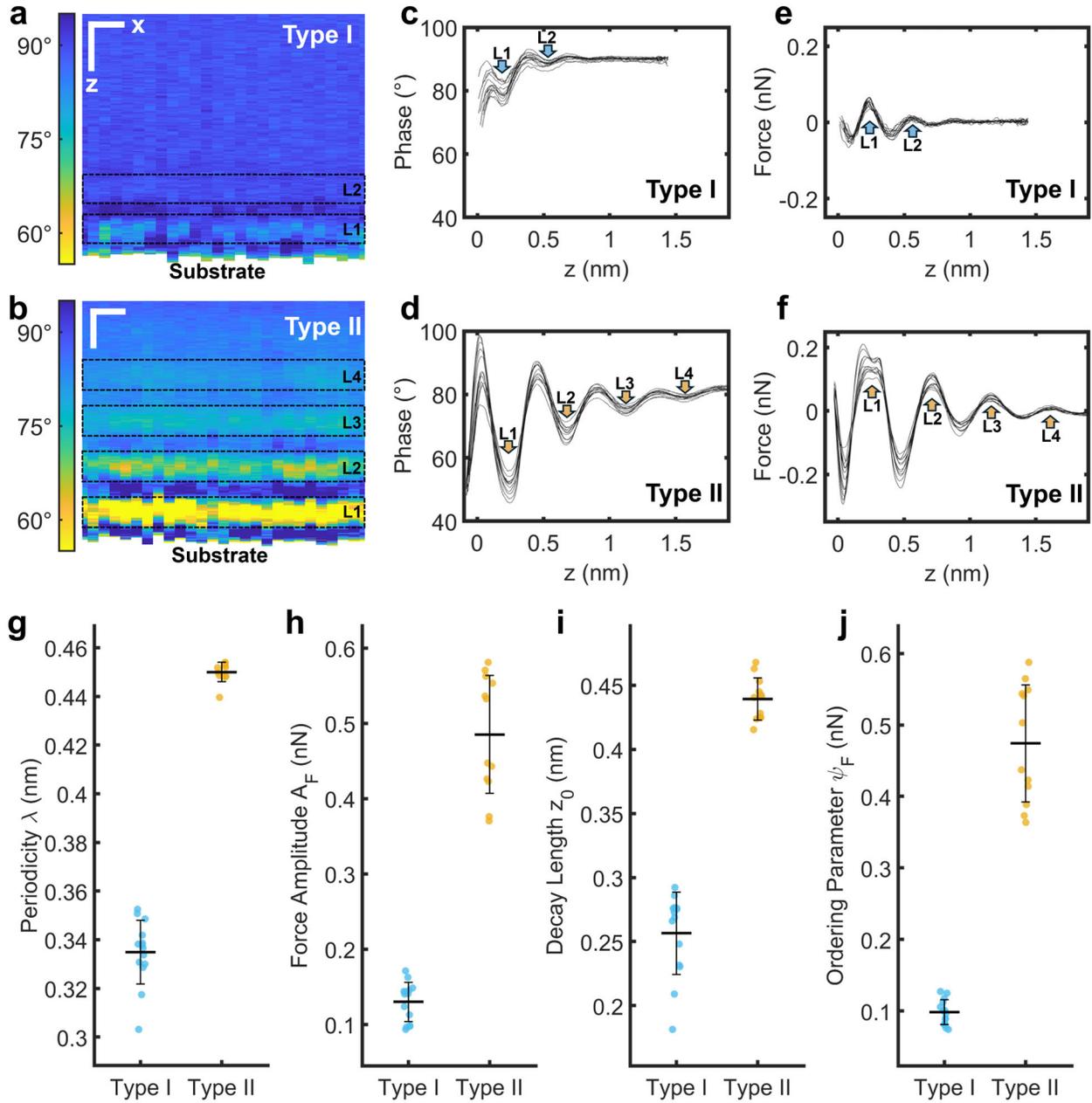

**Figure 2**. 3D-AFM visualization and quantification of interfacial liquid structure. (a,b) x-z phase maps of interfacial liquids above (a) type I area and (b) type II area. Black dashed boxes represent the liquid layers, with layer numbers labeled on the right. The horizontal and vertical scale bars represent 10 Å and 4 Å, respectively. (c,d) Phase vs z curves for (c) type I and (d) type II interfaces, where the z locations of liquid layers are labeled. (e,f) Reconstructed force curves above (e) type I and (f) type II surfaces, where the z locations of liquid layers are labeled. (g–j) Quantification of the oscillatory decay profiles of force curves above a series of type I and type II areas, including (g) periodicity $\lambda$, (h) force amplitude $A_F$, (i) decay length $z_0$, and (j) ordering parameter $\psi_F$. In (g–j), the middle bars correspond to the mean values while the top and bottom caps specify the standard deviation of the data points.

We further reconstructed the conservative part of AFM tip–sample force using the measured cantilever oscillation amplitude and phase. A bi-exponential background force due to long-range interactions was then removed from each force curve.[36,57–59] The background-removed force, $\Delta F$, is directly related to the interfacial liquid density distribution.[20,33,34] $\Delta F$ curves are shown in Figure 2e,f for type I and II domains, respectively. Lower phase generally corresponds to higher $\Delta F$, which roughly corresponds to a larger gradient of the local liquid density.[20] Following existing conventions,[20,32–34,66] we marked the local minima of phase or local maxima of force as liquid layers in this work (Figure 2c–f).

The $\Delta F$ curves from both types of areas exhibited oscillatory decay patterns. Following the predictions of classical liquid theory, we assumed an approximate formula for the force-distance relation:[20,67–69]

$$\Delta F(z) = A_F cos(qz - \varphi)e^{-z/z_0}, \tag{1}$$

where $A_F$, $\lambda = 2\pi/q$, $\varphi$, and $z_0$ are the force amplitude, oscillation periodicity, phase factor, and decay length of the damped oscillation function, respectively. It is known that Eq. 1 is only valid beyond the first liquid layer, since it does not include the specific interactions between the first layer of liquid and the substrate.[20] Therefore, to obtain reasonable parameters, we only used data at z > 0.2 nm and z > 0.68 nm for the fitting of type I and type II curves, respectively. Figure 2g–i shows the statistical comparison of three important parameters—$\lambda$, $A_F$, and $z_0$— between type I and II surfaces. The parameter $\varphi$ is sensitive to the z offsets. After applying a manual offset to align the force curves, $\varphi$ is no longer a valid descriptor of the interfacial liquid structure and will not be further discussed in this work.

The density oscillation period $\lambda$, also known as the interlayer distance, represents the average intermolecular distances of local liquid molecules and serves as a key metric for differentiating molecular species.[32–34] Previous measurements, using 3D-AFM or X-ray scattering, revealed a $\lambda$ value of ~0.33 nm for water,[32–34,36,49,70,71] ~0.4 nm for ethylene carbonate/dimethyl carbonate,[72] ~0.5 nm for n-octane and n-pentadecane,[32,34] and ~0.6 nm for tetraglyme.[73] For a given interfacial liquid species, $\lambda$ tends to be same as the average intermolecular distance of the corresponding bulk liquid, and is largely independent of the substrate composition.[34,67,69,74] Figure 2g shows that $\lambda$ for type I areas is 0.33±0.013 nm, consistent with previously reported interlayer distance of water.[32–34,36,49,70,71] In contrast, we found that $\lambda = 0.45\pm0.004$ nm for type II regions, larger than that expected for water, suggesting a different local liquid composition.

The force amplitude $A_F$ was found to be 0.13±0.026 nN and 0.48±0.078 nN for type I and type II areas, respectively (Figure 2h). In general, $A_F$ reflects the combined effects of intermolecular forces within the liquid and interactions between the liquid and the substrate.[20,50,53,75] At the Pt–water interface, our observation of a significantly larger $A_F$ at type II regions indicates two possibilities: 1) the local intermolecular interaction among type II liquid species is stronger than that of water; and 2) the type II liquid has a stronger affinity/attraction to the Pt surface compared to water. Either of these two scenarios would lead to a more rigid (less mobile) interfacial liquid. It is noteworthy that a tiny notch appears in the first force peak (L1) of type II liquid (Figure 2f). This peak splitting implies that the L1 layer may contain both a weakly bound or physisorbed sub-layer (first sub-peak) and a nearby mobile, liquid sub-layer (second sub-peak). The possible physisorption effect is consistent with a stronger Pt–liquid interaction that may be (partially) responsible for the larger $A_F$ at type II regions (compared to that in type I domains).

Another metric for interfacial liquid structure, decay length ($z_0$), was rarely discussed in previous literature. $z_0$ characterizes the length scale over which the interfacial liquid density oscillation gradually decays, ultimately merging with the homogeneous bulk density distribution. Previous density functional analysis has shown that $z_0$ is mainly governed by intermolecular correlations within the liquid, with minimal influence from liquid–substrate interactions.[34,74] Figure 2i shows that the decay length for type I areas is 0.26±0.030 nm, smaller than 0.44±0.016 nm for type II areas. This means that type I liquid transitions into isotropic behavior at shorter z distances than type II.

The $\lambda$, $A_F$, and $z_0$ parameters in type I regions all match with previously reported values for interfacial water.[32–34,36,49,70,71] On the contrary, these three parameters for type II areas are close to those of interfacial n-alkanes (e.g., hexane, octane, and pentadecane) and adventitious hydrocarbon-contaminated hydrophobic substrate–water interfaces shown in prior studies.[32–34]

The layering or density oscillations of interfacial liquids reflect their higher level of ordering compared to the bulk counterparts. Previous work used scaled decay length ($z_0/\lambda$) as a metric for the intermolecular correlation effect.[34] However, this parameter is primarily dependent on the properties of the corresponding bulk liquid, and cannot fully capture the interfacial liquid ordering. To more comprehensively quantify the ordering level of interfacial liquids, here we propose a new descriptor, the ordering parameter $\psi_F$, calculated as the product of the force amplitude and the ratio of decay length to oscillation periodicity:

$$\psi_F = A_F \cdot \frac{z_0}{\lambda}. \tag{2}$$

Here, larger force amplitude indicates more pronounced density oscillation, and larger decay length reflects stronger spatial correlation along the z direction. Normalizing decay length by periodicity accounts for the molecular size effect. Thus, a higher $\psi_F$ value reflects a more ordered interfacial liquid structure. As shown in Figure 2j, $\psi_F$ was determined to be 0.098±0.018 nN and 0.47±0.082 nN for type I and type II domains, respectively. This suggests a markedly higher degree of ordering in type II interfacial liquids. The ordering level difference can be more intuitively perceived in phase maps and phase/force profiles, where only two layers can be clearly observed in type I areas, whereas at least four layers in type II regions have been detected by 3D-AFM (Figure 2a–f).

## 3.3. Bulk Liquid Structure and Their Correlation to Pt–Water Interfaces

Since $\lambda$ of an interfacial liquid is known to be the same as the intermolecular distance of the corresponding bulk liquid,[34,67,69,74] we proceeded to measure a series of bulk liquid structure, with the goal of identifying the possible compositions of the liquid layers at Pt–water interfaces. To this end, we conducted WAXS measurements of a host of bulk liquid candidates, including pure water and several pure n-alkane liquids (n-heptane, n-octane, n-nonane, n-decane, n-undecane, n-dodecane, n-tridecane, and n-tetradecane). Results for individual samples before and after background-subtraction are shown in Figure S7.

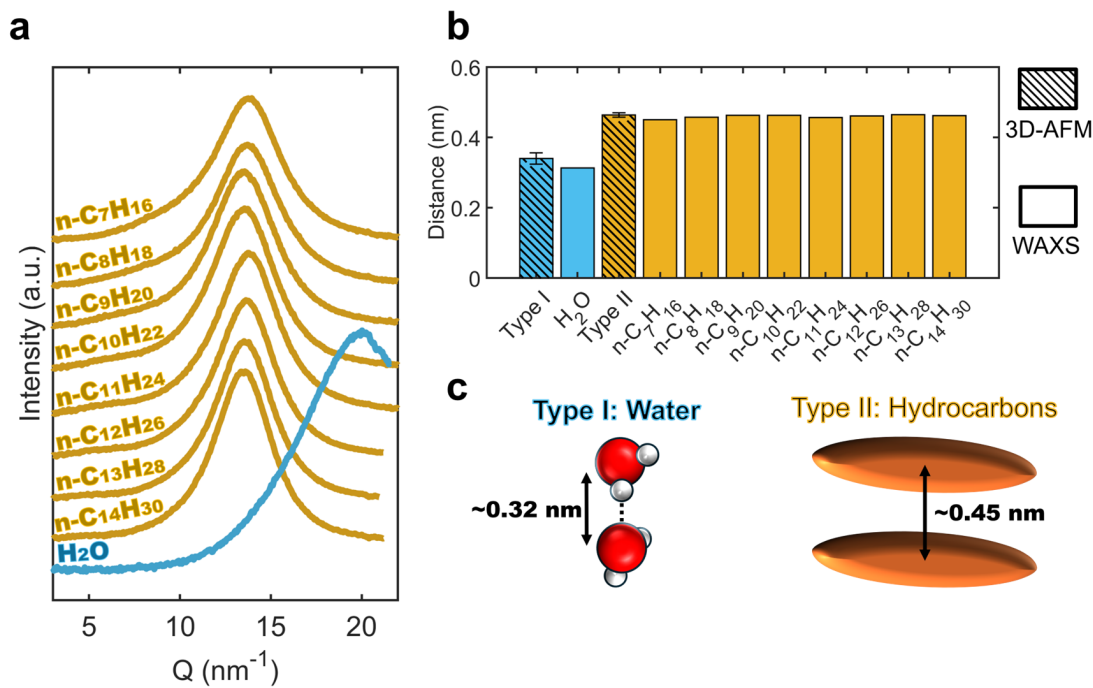

**Figure 3**. Correlation between bulk and interfacial intermolecular distance. (a) WAXS intensity vs Q results of water and a series of n-alkanes (n-$C_7H_{16}$ – n-$C_{14}H_{30}$). (b) Comparison of interlayer distances obtained from 3D-AFM at Pt–water interface and intermolecular distances from WAXS of bulk liquids. Error bars on the 3D-AFM columns represent standard deviation. (c) Schematics of the intermolecular configurations of water and n-alkanes both in bulk and at the interface.

Figure 3a summarizes the background-subtracted scattering intensity of all the measured bulk liquids vs Q. The scattering pattern is a reciprocal space representation of the real-space pair correlation function (PCF). A scattering peak at a Q space position $Q_{peak}$ corresponds to a peak in the real-space PCF at an intermolecular distance of $2\pi/Q_{peak}$. For all liquid samples measured, only one scattering peak was observed within the measured Q range, indicating that the statistical average of liquid structure only has one prevailing nearest neighbor intermolecular distance. We extracted the peak position of the measured bulk liquids and obtained the prevailing real-space intermolecular distance via $2\pi/Q_{peak}$ (Figure 3b). The intermolecular distance for bulk water was observed to be 0.31 nm, which matches existing standard results and is determined by the intermolecular hydrogen bond (Figure 3c).[71,76] For all n-alkane species, the intermolecular distances are nearly the same, with an average value of 0.46±0.0048 nm. The chain length-independence of the n-alkane liquids was demonstrated in previous reports,[77,78] and was attributed to the constant intermolecular separation in the direction normal to the carbon chain (Figure 3c). This does not contradict the dynamic, mobile nature of the liquid molecules. The nearest neighbors of molecules can have a much higher probability of side-by-side packing compared to end-to-end packing at any given moment, and the time-averaged intermolecular distance reflects solely the side-to-side distance.

Comparing the WAXS and 3D-AFM results, we observed good match between the interlayer spacing of type I interfacial liquid and the intermolecular distance in bulk water, and between type II interfacial spacing and the intermolecular distances of n-alkanes (Figure 3b). Thus, we conclude that the dominant species at type I and type II interfacial domains are most likely water and hydrocarbon molecules, respectively. Although the exact chemical nature of the hydrocarbon species is still unknown, they likely also consist of carbon chains as the backbone, similar to n-alkanes. The side-by-side packing of these carbon chains is expected to govern their intermolecular distance.

### 3.4. Overall 3D Microscopic Structure of Pt–Water Interfaces and Implications

Combining the results shown above, we summarize the microscopic structure of Pt–water interface in Figure 4. Overall, we observed the co-existence of two types of 3D domains. Type I domains are characterized by direct contact between water molecules and the substrate, exhibiting hydrophilic behavior. In contrast, type II domains feature hydrophobic behavior, with hydrocarbons both covering the Pt surface and extending ~2 nm from substrate surface.

The observed heterogeneity and associated interfacial liquid structure arise from the dynamic interplay among substrate, water, and hydrocarbons. Hydrocarbon species are known to be ubiquitous in ambient environments and may originate from various sources such as ambient air, gas emissions, and even biological activities like breathing.[79–83] Several possible pathways can introduce hydrocarbons to the Pt–water interface. Since no type II features were observed during imaging in air (Figure 1a), one plausible route is that trace amount of airborne hydrocarbon molecules dissolved into water and then aggregated on the Pt surface. The feasibility of this process can be inferred from a previous computational study, which demonstrated a solvation free energy profile of Pt(111)–water interface that favors the adsorption of hydrophobes.[84] Additionally, hydrocarbons may also be liquid-born, originating from the trace contaminants of the liquid containers or the Milli-Q water source.

When a clean Pt is immersed in water, it is possible that small, spatially scattered hydrocarbon nuclei initially form on Pt via aforementioned routes. These nuclei increase the local affinity for additional hydrophobes in water, thereby accelerating the local hydrocarbon aggregation, resulting both lateral extension of the stripy surface domains and the vertical accumulation of hydrocarbon liquid layers. Notably, the observed damped oscillation profiles of the local hydrocarbon layers (Figure 2f) closely resemble those of pure bulk alkanes at solid surfaces,[32,34] even though the surrounding bulk liquid is water in this study (Figure 4). This indicates that the interfacial hydrocarbon liquids are likely phase segregated from the surrounding bulk water, similar to the liquid–liquid phase separation effects occurring in many systems ranging from oil/water mixtures to biomolecular condensates.[85–88]

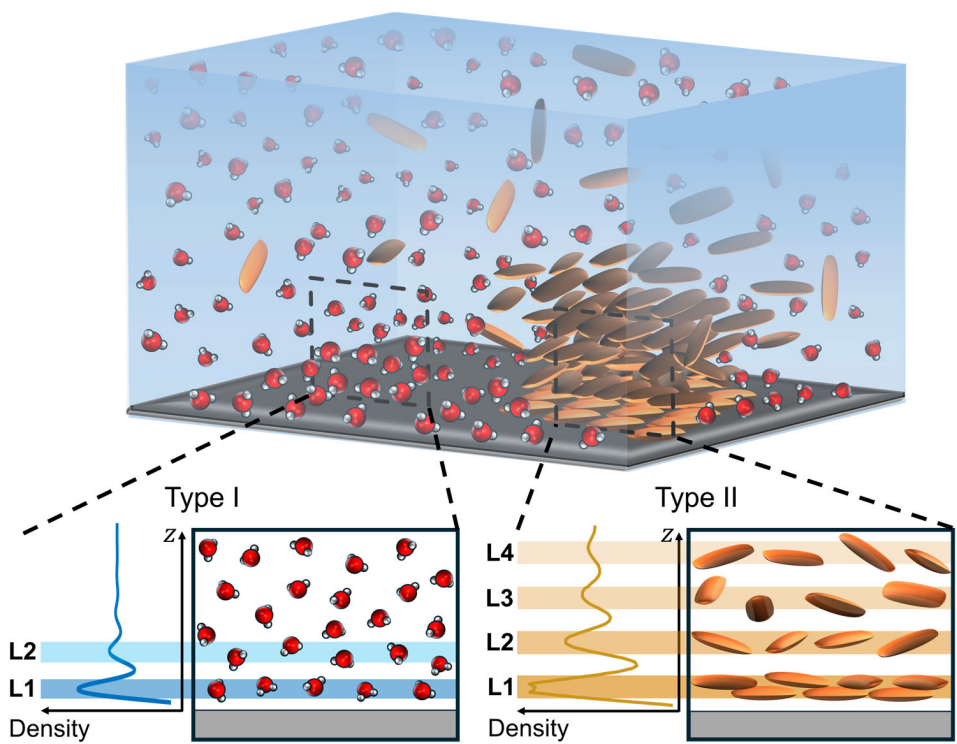

**Figure 4.** Schematics of Pt–water interfacial structure. Some areas of the Pt surface display hydrophilic behavior (type I), where water-dominated interfacial liquid is less ordered with smaller interlayer distance and decay length. Other areas, in contrast, display hydrophobic behavior (type II) and are covered with hydrocarbon contaminations that exhibit stripy surface morphology. The interfacial liquids are also dominated by hydrocarbons extending ~2 nm above the substrate, exhibiting oscillatory decay pattern with larger periodicity and longer decay length, forming an overall more ordered structure.

Our findings can have strong implications for the electrochemical processes that hinge on Pt–aqueous solution interfaces. It has been reported that certain added hydrophobic species, once accumulated at the Pt(111)–aqueous solution interface, can enhance ORR by preventing the adsorption of poisoning species.[45,89] Our results reveal that, even without intentionally added external molecular species, adventitious hydrocarbons may still accumulate at local sites of the Pt surface. These local hydrocarbons may either enhance or suppress the local electrochemical reactions, depending on the strength of Pt–hydrocarbon interaction and the nature of the reaction process. Additionally, these local hydrophobes might dynamically reconfigure during electrochemical cycling, similar to our previous observations on graphite–water interfaces.[36] Therefore, a comprehensive understanding or control of electrochemical reactions will require the knowledge on the real-time configuration of the interfacial distribution of hydrocarbons. Our results provide a first step for the in-situ observation of 3D interfacial configurations of Pt–water interfaces. Future investigations on electrochemical responses of the observed interfacial structures will offer additional insights into the reaction dynamics.

## 4. Conclusions

To conclude, we employed 3D-AFM to probe Pt–water interfaces and discovered the coexistence of flat and stripe-like domains on the surface. On both domains, the interfacial liquid exhibited damped oscillation force profiles. We quantified the key metrics of these profiles, including oscillation periodicity $\lambda$, force amplitude $A_F$, decay length $z_0$, and a derived ordering parameter $\psi_F$. Stripe-like domains (type II) have stronger force oscillation, longer periodicity and extended decay length than flat domains (type I), indicating a more ordered liquid structure. Through correlative comparison of intermolecular distances measured via 3D-AFM and WAXS experiments, we conclude that type I and II interfaces are dominated by water and adventitious hydrocarbons, respectively. These findings have strong implications for the understanding and design of metal–water interfaces for many electrochemical and catalysis applications.

**Associated Content**

**Supporting Information**

The Supporting Information is available free of charge online:

Key AFM setup and cantilever parameters (Table S1); imaging parameters used to acquire all AFM results (Table S2). Verification of the Pt surface structure (Figure S1); typical surface morphologies of type I and type II surface areas of Pt immersed in water (Figure S2); surface morphology of the same area on Pt immersed in water from continuous AFM topographic scans (Figure S3); a gallery of x-z phase maps of type I and II interfaces (Figures S4–S6); WAXS intensity vs Q of empty tube background, tube and sample together, and sample alone from n-heptane, n-octane, n-nonane, n-decane, n-undecane, n-dodecane, n-tridecane, and n-tetradecane, and water (Figure S7) (PDF)

**Notes**

The authors declare no competing financial interest.

**Acknowledgements**

We acknowledge the support from the Beckman Young Investigator Award provided by the Arnold and Mabel Beckman Foundation and the Sloan Research Fellowship from the Alfred P. Sloan Foundation. The experiments were performed in part in the Carl R. Woese Institute for Genomic Biology and in the Materials Research Laboratory at the University of Illinois Urbana-Champaign.